\documentclass[pss]{wiley2sp} 
\usepackage{amsmath}

\begin{document}

\title{Quantum Criticality and Novel Phases: Summary and Outlook}

\titlerunning{QCNP09: Summary and outlook}

\author{%
  A. J. Schofield
}
\authorrunning{A. J. Schofield}

\mail{e-mail
  \textsf{ajs@th.ph.bham.ac.uk}, Phone: +44-121-4144671}

\institute{%
  School of Physics and Astronomy, \\
  University of Birmingham, \\
  Edgbaston, Birmingham, B15 2TT, \\
  United Kingdom.
}

\received{XXXX, revised XXXX, accepted XXXX} 
\published{XXXX} 


\abstract{%
%
%
%
\abstcol{
This conference summary and outlook provides a personal
overview of the topics and themes of the August 2009 Dresden meeting
on quantum criticality and novel phases. The dichotomy between the
local moment and the itinerant views of magnetism is revisited and
refreshed in new materials, new probes and new theoretical ideas.  New
universality and apparent zero temperature phases of matter move us
beyond the old ideas of quantum criticality. This is accompanied by
alternative pairing interactions and as yet unidentified phases
developing in the vicinity of quantum critical points.}  
{In discussing novel order, the magnetic analogues of
superconductivity are considered as candidate states for the hidden
order that sometimes develops in the vicinity of quantum critical
points in metallic systems. These analogues can be thought of as
``pairing'' in the particle-hole channel and are tabulated.  This
analogy is used to outline a framework to study the relation between
ferromagnetic fluctuations and the propensity of a metal to nematic
type phases---which at weak coupling correspond to Pomeranchuk
instabilities. This question can be related to the fundamental
relations of Fermi liquid theory.}}

%
%

\maketitle   

\section{Introduction}

The vibrant growth of research activity in the area of quantum
criticality is amply illustrated by the presentations at the Dresden
meeting on Quantum Criticality and Novel Phases in August 2009 and
this associated proceedings. It is a subject area which is well-served
by reviews, both technical~\cite{lohneysen_2006a} and
non-specialist~\cite{sondhi_1997a,coleman_2005a}, as well as
textbooks~\cite{sachdev_1999a} and this brief conference summary is
certainly not intended to compete with these nor even cover all the
activity of the Dresden meeting. Rather it gives a subjective pathway
through some of the main themes of the meeting and some personal
thoughts on key questions outstanding in the field.

\section{Quantum criticality}

Quantum criticality has an astonishing legacy which, in the context of
itinerant systems, can be traced back to the 1960s with the puzzling
behaviour of $d$-metals at or on the border of ferromagnetism.
Materials like palladium, $\rm Ni_3Al$, $\rm Ni_3Ga$, $\rm YNi_3$ and
$ \rm ZrZn_2$ combined facets of two distinct views of magnetic
behaviour: local moment magnetism and itinerant moment (Stoner)
magnetism.  The small ordered moment and metallic characteristics
which point to the latter picture stand at odds with their high
temperature Curie-Weiss susceptibilities and large fluctuating moment
more characteristic of a local moment system.  The resolution of this
paradox is the inclusion of critical spin-fluctuations which
renormalize the magnetic equation of
state~\cite{lonzarich_1985a}. Numerous researchers have played
significant roles in understanding this---the ``conventional theory''
of itinerant quantum criticality---culminating in the renormalization
group treatment of the Hertz-Millis
action~\cite{hertz_1976a,millis_1993a}
\begin{equation}
S= \! \! \int \! \sum_{\omega_n} d^D q \left[
\left(x-x_c\right) + q^2 + \frac{|\omega_n|}{q^{z-2}} |\psi|^2 + u |\psi|^4
  \cdots \right].
\end{equation}
As we have seen in this meeting, there are many cases where this
theory seems to give both a qualitative (for example, the work
presented here by Shinsaku Kambe {\it et al.}~\cite{kambe-qcnp09} on
NMR of $\rm USn_3$) and sometimes a quantitative agreement with
experiment (as shown by Gilbert
Lonzarich~\cite{smith_2008a,lonzarich-qcnp09}).

In the latter case we saw the ferromagnetic metal demonstrating the
same physics as the Reizer singularity which arises from the weak
current-current interactions in all metals~\cite{reizer_1989a}.  These
unscreened long-range interactions would ultimately drive any metal
into a non-Fermi liquid state---albeit at micro-Kelvin
temperatures. The idea that one might employ this mechanism and
amplify it to make such a state visible to experiment once inspired
ideas like the $U(1)$ gauge theory descriptions of the resonating
valence bond (RVB) state.  These were conjectured for the high-$T_c$
cuprates to account for, among other things, the linear $T$ normal
state resistivity~\cite{ioffe_1990a,wheatley_1992b}. Yet, as the
ferromagnetic metal illustrates, quantum criticality can also generate
long-ranged, unscreened interactions and associated power-law
resistivities. This has led many to see the cuprates as quantum
critical systems. The precise nature and location of the putative
quantum critical point in the cuprates is hotly
debated~\cite{anderson_2004a}. Fresh insight by Subir Sachdev
presented at this meeting~\cite{sachdev-qcnp09} addresses this by
showing that superconductivity and spin-density wave order interfere
to push the associated quantum critical point into the underdoped
region.

\begin{figure}[t]%
\includegraphics*[width=\linewidth]{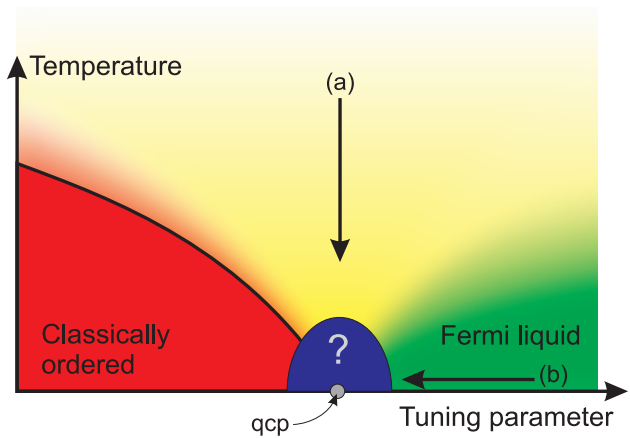}
\caption{%
Schematic phase diagram of a quantum critical metal with a
novel phase (labelled ``?'') centred over the quantum critical
point (qcp). The power-law quantum critical metal is approached
along path (a). Path (b) allows the instability to the novel
phase to be studied as a Fermi liquid theory instability.}
\label{qcp}
\end{figure}

Nevertheless, there remains the more general question of whether
power-law temperature dependencies of scattering down to the lowest
temperatures, often now presented in the form of colour coded
scattering maps, are always indicative of an associated quantum
critical point (Fig.~\ref{qcp}). Obvious counter examples, such as the
Luttinger liquid state in 1D, were discussed at the meeting by Thierry
Giamarchi. So too were the Griffiths singularities in disordered
quantum magnets which again can give singular critical behaviour for a
range of parameters as presented by Thomas
Vojta~\cite{vojta-qcnp09}. It seems unlikely that this exhausts the
list given the experiments in MnSi at pressures above the quantum
phase transition which suggest a critical state of
matter~\cite{doiron-leyraud_2003a}.

Despite the successes of the ``conventional theory'' of the itinerant
quantum critical metal, a growing set of counter examples has emerged.
They seem to lie outside the theory at a fundamental level as typified
by two systems.  ${\rm CeCu_{6-x}Au_x}$, though nominally above the
conventional theory's upper critical dimension, shows $E/T$ scaling
and anomalous exponents, as well as evidence of critical properties
throughout the Brillioun zone~\cite{schroder_2000a}. ${\rm
YbRh_2Si_2}$ shows evidence of multiple energy
scales~\cite{gegenwart_2007a} converging to zero at its
magnetic-field-driven quantum-critical point. New ``bad actors'' have
been presented here including ${\rm Ce_3Pd_{20}Si_{6}}$ by Silke
Paschen~\cite{paschen-qcnp09}. One common denominator amongst them is
that these materials are heavy fermion systems where the quantum
criticality emerges from a Fermi liquid state which is already exotic.

Like the $d$-metals where our story began, these $f$-metal materials
combine local moments and itinerancy, but unlike them, they do this
at microscopic level with distinct local moments which interact weakly
with separate conduction bands. Yet despite this microscopic
separation apparent at high energies, these degrees of freedom combine
at low energy. Together they form a metallic state comprised of
quasiparticles which reveal their local moment constituents
entropically in their heavy mass and numerically in a Fermi volume
that includes the magnetic moments as if they were itinerant.  The
suggestion that a new form of quantum criticality associated with
breakup of these composite quasiparticles and revealed in the collapse
of the large-volume Fermi surface~\cite{coleman_2001a} was emphasized
by Qimiao Si~\cite{si-qcnp09} and recurred in multiple presentations
here.

The definitive proof of a volume change in the Fermi surface at a
quantum critical point is arguably still lacking. Quantum oscillation
experiments have been the pre-eminent technique for the pinning down
the fermiology of correlated electron systems. In that vein, the work
of Shishido {\it et al.}~\cite{shishido_2005} in ${\rm CeRhIn_5}$ is
very promising and so too is the research presented here by Stephen
Julian~\cite{sutton-qcnp09} beginning the elucidation of the Fermi
surface of ${\rm YbRh_2Si_2}$ with the same technique. In this
conference we were shown two other techniques which have been used to
great effect in correlated oxide materials but are now being applied
to the heavy fermion systems. Photoemission spectroscopy offers the
prospect of imaging the electron structure of quasiparticles and we
saw, in its angle-integrated form, the emergence of the Kondo
resonance~\cite{klein_2008a}. When angle-resolved, one might hope to
see the Fermi surface itself~\cite{vyalikh_2009a}. Of course observing
the putative Fermi surface volume change at a quantum critical point
is a significant challenge for ARPES. The second technique which has
been employed to great effect in the correlated oxides is low
temperature STM and at this conference we saw the results of its
application to heavy fermion systems by Seamus Davis and his
group. We were shown the STM view of how local moments and
conduction electrons combine to form composites and many were struck
by how much more insight this new experimental development might soon
give us.

One should not overlook the role that transport measurements,
particularly magneto-transport, can make in characterizing the nature
of the quantum critical point~\cite{coleman_2005b}. The appearance of
a linear-in-magnetic-field transverse magnetoresistance should
characterize a conventional density wave
transition~\cite{fenton_2005a}.  This has now been observed
experimentally in ${\rm Ca_3Ru_2O_7}$ by Kikugawa {\it et
al.}~\cite{kikugawa_2010a}. The complexity of density wave and other
transitions in this material was presented to us by Malte
Grosche~\cite{grosche-qcnp09}. Other work presented in this conference
by Friedemann and co-workers argues that evidence for $E/T$ scaling
can be seen in the Hall effect of ${\rm YhRh_2Si_2}$, thereby linking
for the first time the anomalous properties of this material with
the other canonical anomalous quantum critical system, ${\rm
CeCu_{6-x}Au_x}$.

The theoretical ideas to account for the rich behaviour of metallic
quantum critical matter are far from settled and experiments may even
suggest that they are still too conservative. This is not just the
apparent phase of non-Fermi liquid matter in MnSi, but the observation
that the two collapsing energy scales seen in ${\rm YbRh_2Si_2}$ do
not need to hit zero at the same point~\cite{friedemann_2009a} but
could open a spin-liquid window between the classically-ordered and
quantum-disordered phases. 

The inspiration for new theories will not just come from heavy fermion
experiments. In this meeting we saw two other classes of quantum
system, each of which provide a new window into correlated quantum
behaviour. Quantum magnetism offers systems which are often
conceptually simpler than their itinerant counterparts (without the
damping effects of other low lying degrees of freedom) and has a long
history of inspiring new theoretical ideas from
RVB~\cite{anderson_1973d,anderson_2004a} to deconfined
criticality~\cite{senthil_2004a}. Giamarchi stressed the utility of
quantum spin systems as simulators of Bose
condensates~\cite{giamarchi_2008a} as well as providing detailed tests
of one of the few non-Fermi liquid phases we think we understand---the
Luttinger liquid. In fact two distinct recent experiments suggest
Luttinger liquid properties may extend over a wider range than naively
thought~\cite{ruegg_2008a,jompol_2009a}.

The other more recent quantum playground comes in the form of trapped
ultra-cold atomic gases held in optical lattices. This rapidly
developing field was surveyed for us by Immanuel Bloch and Achim
Rosch~\cite{bloch-qcnp09}. The present state of the art experiments in
Fermi gases go down to temperatures that are still about 15\% of the
Fermi temperature and, in the Mott insulating region, are a factor of
two or so higher than the expected Ne\'el temperature. Thus we are not
yet at a stage where we can simulate directly the quantum critical
systems being studied in the solid state. Nevertheless the prospect of
being able to address single atomic sites means that when we reach the
appropriate regimes we can expect true quantum simulation of many of
the models whose physics continues to elude theoretical
description. Yet, as Achim Rosch pointed out in a complementary theory
talk, not only can experiments on the Mott region be well matched to
dynamic mean-field theory predictions~\cite{schneider_2008a}, but
exploiting the metastability intrinsic to cold-atomic gases allows a
novel form of pairing in the repulsive Hubbard
model~\cite{rosch_2008a}. This naturally links to the second key theme
of the meeting.

\begin{figure}[t]%
\includegraphics*[width=0.45\linewidth]{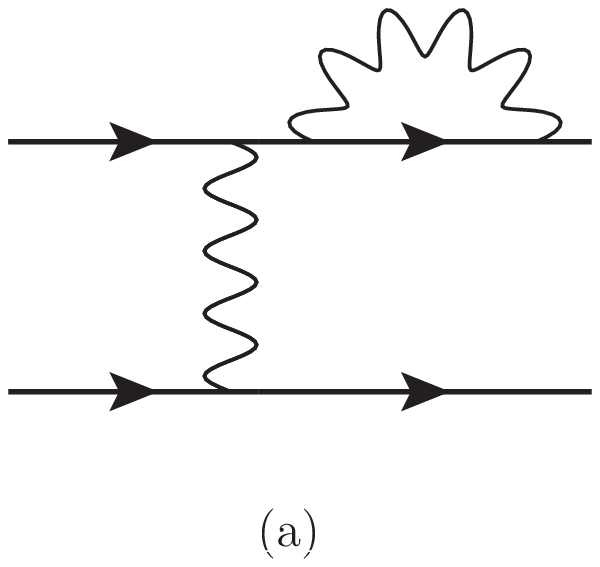}
\hfill
\includegraphics*[width=0.45\linewidth]{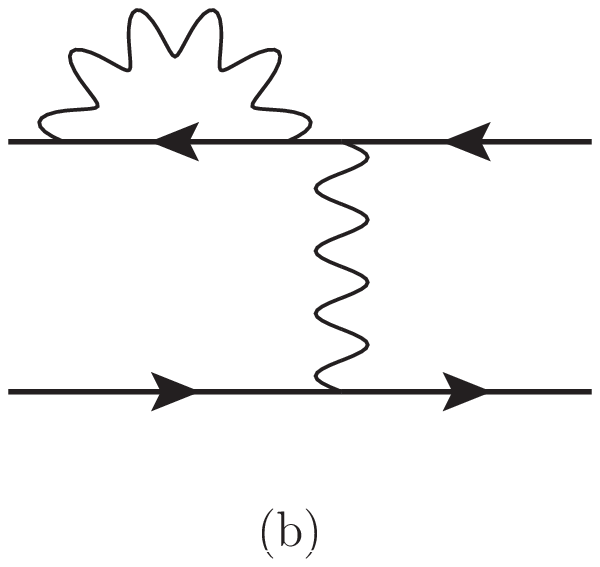}
\caption{%
(a) A schematic of the lowest order spin-fluctuation interactions
in the particle-particle channel responsible for pairing-making and
pair-breaking. (b) Rotating the direction of the propagator indicates
that these same interactions could ``pair'' in the particle-hole
channel. A singularity here would denote an instability to a form of 
magnetism (or charge order).}
\label{pairing}
\end{figure}

\section{Novel Phases}

One of the incentives for studying quantum critical matter
is that at or near the quantum critical point we often see the
appearance of new ordered states (shown schematically in
Fig.~\ref{qcp}). The natural explanation is that the fluctuations
induced by the quantum melting of an ordered phase can drive other
instabilities. The classic example is the appearance of
superconductivity near antiferromagnetic quantum critical
points~\cite{mathur_1998a,monthoux_2007a}.

\begin{table*}[t]
  \caption{Unconventional superconductors where the Cooper pairs have
internal angular momentum and/or centre of mass momentum or a spacial
texture have exotic magnetic analogues in the particle-hole channel.
The analogy is explored here where $V$ is the pairing interaction and
$g$ denotes a symmetry factor which can depend on spin and orientation
around the Fermi surface.
}
  \begin{tabular}[htbp]{@{}|c|c|@{}}
    \hline
    Superconductor (particle-particle channel) & Magnet (particle-hole
    channel) \\
    \hline
    Conventional $s$-wave  & Stoner ferromagnet  \\
    $\displaystyle \Delta^*= \sum_{k',\sigma} V \langle
    \hat{c}^\dagger_{k'\sigma} \hat{c}^\dagger_{-k'\sigma}\rangle$  &
    $\displaystyle M_\sigma=\sum_{k',\sigma'} g_{\sigma,\sigma'} \langle
    \hat{c}^\dagger_{k'\sigma'}\hat{c}^{}_{k'\sigma'}\rangle$ \\
    \hline
    Unconventional ($d$-wave etc.) & Pomeranchuk ($d$-wave nematic~\cite{oganesyan_2001b} etc.) \\
    $\displaystyle \Delta^*(k) = \sum_{k'\sigma} V_{kk'} \langle
    \hat{c}^\dagger_{k'\sigma} \hat{c}^\dagger_{-k'\sigma}\rangle$  &
    $\displaystyle M_\sigma(k)=\sum_{k',\sigma'} g_{k,k';\sigma,\sigma'} \langle
    \hat{c}^\dagger_{k'\sigma'}\hat{c}^{}_{k'\sigma'}\rangle$ \\
    \hline
    Inhomogeneous ($s$-wave FFLO~\cite{fulde_1964a,larkin_1964a}) & Inhomogeneous (spiral, density-wave) \\
    $\displaystyle \Delta^*(q) = \sum_{k'\sigma} V \langle
    \hat{c}^\dagger_{k'+q/2,\sigma} \hat{c}^\dagger_{-k'+q/2,\sigma}\rangle$  &
    $\displaystyle M_\sigma(q)=\sum_{k',\sigma'} g_{\sigma,\sigma'} \langle
    \hat{c}^\dagger_{k'+q/2,\sigma'}\hat{c}^{}_{k'+q/2,\sigma'}\rangle$ \\
    \hline
    Inhomogeneous ($d$-wave FFLO etc.) & Inhomogeneous
    ($d$-density-wave~\cite{chakravarty_2001b} etc.) \\
    $\displaystyle \Delta^*(k,q) = \sum_{k'\sigma} V_{kk'} \langle
    \hat{c}^\dagger_{k'+q/2,\sigma} \hat{c}^\dagger_{-k'+q/2,\sigma}\rangle$  &
    $\displaystyle M_\sigma(k,q)=\sum_{k',\sigma'} g_{k,k';\sigma,\sigma'} \langle
    \hat{c}^\dagger_{k'+q/2,\sigma'}\hat{c}^{}_{k'-q/2,\sigma'}\rangle$ \\
    \hline
    Abrikosov flux lattice & Skyrmion spin-texture~\cite{bogdanov_1989a} \\
    \hline
  \end{tabular}
  \label{analogue}
\end{table*}

Spin-fluctuation mediated superconductivity and superfluidity has,
like quantum criticality itself, been much studied theoretically and
experimentally from the 1960s onwards. Condensing this wealth of
knowledge into a single set of principles is challenging yet, in
discussion at the meeting, Gil Lonzarich articulated what might be
termed ``Lonzarich's
Rules''~\cite{monthoux_1999a,monthoux_2001a,monthoux_2002a} in homage
to the older ``Matthias' Rules''\cite{gaule_1963a} which inspired a
previous generation's search for phonon mediated superconductors. Like
``Matthias' Rules'' these have not explicitly appeared in print but are
otherwise unlike them in almost every respect. So in looking for
higher superconducting transition temperatures one should:
\begin{itemize}
\item Look on the border of magnetism ({\it i.e.} near a quantum
critical point),
\item Prefer antiferromagnetism to ferromagnetism,
\item If ferromagnetic, favour uniaxial anisotropy,
\item Prefer 2D electronic structures over 3D,
\item Look for materials with a large spin-fluctuation scale,
\item Prefer single band materials or nested multiband materials,
\item Avoid competing antiferromagnetic and ferromagnetic fluctuations.
\end{itemize}
The question of how the order parameters of magnetism and
superconductivity can compete and even coexist was posed by Georg
Knebel~\cite{knebel-qcnp09}.  In essence these ``rules'' attempt to
capture that competition between the Cooper pair-making and Cooper
pair-breaking effects of the spin fluctuations [see
Fig.~\ref{pairing}(a)] and how the resulting gap structure interplays
with underlying electron structure to optimize the condensation
energy.  Each of these rules come with caveats and counter examples
yet new superconductors which seem to conform to these guidelines were
presented here such as ${\rm Ce_2PdIn_8}$~\cite{kaczorowski_2009a}.
Pascoal Pagliuso also stressed the role of low dimensionality and
hybridization as key ingredients in the quest for new superconducting
materials.  Moreover Satoru Nakatsuji has discovered superconductivity
in the hole system ${\rm
\beta-YbAlB_4}$~\cite{nakatsuji_2008a,nakatsuji-qcnp09} which points
all the more to a universal mechanism for superconductivity again from
a novel metallic state.

However, the conference also saw a range of superconductors which were
suggestive of other forms of pairing interaction. If a valence
transition could be made critical then Shinji Watanabe argued it could
cause pairing~\cite{watanabe-qcnp09}, as perhaps is seen in ${\rm
CeCu_2(Si_{1-x}Ge_x)_2}$~\cite{yuan_2003a}. The corresponding magnetic
analogue would be a pairing from metamagnetic fluctuations---perhaps
the case in ${\rm UGe_2}$,~\cite{kitaoka_2005a,sandeman_2003a}. Tuson
Park raised the intriguing question as to whether the local character
of some of the anomalous heavy fermion quantum critical points could
also drive superconductivity in, for example, ${\rm
CeRhIn_5}$~\cite{park_2008a,park-qcnp09}. Indeed, given the rich
magnetic phase diagram of materials like ${\rm
CeCoIn_5}$~\cite{kenzelmann_2008a}, determining the dominant driving
mechanism for superconductivity ({\it i.e.} the one that contributes
most to the condensation energy) is an outstanding question. It is a
question that is relevant not just for the heavy fermion systems but
also the iron based superconductors where issues such as local moment
physics versus itinerancy are actively being investigated as we heard
from Nan-Lin Wang~\cite{wang-qcnp09} and from Maw-Kuen
Wu~\cite{wu-qcnp09}.

Superconductivity is not the only novel phase seen near quantum
critical points. The enigmatic 17K transition~\cite{palstra_1985a} in
${\rm URu_2Si_2}$ continues to be a cause of fascination and new ideas
and experiments were described in numerous posters presented here. As
stressed by Seamus Davis in describing STM work on this problem, this
is not ``hidden order''---its presence in the phase diagram is
unmistakable. Rather it is more appropriately ``dark order'' in that
it is transparent to our usual probes, yet new theory and experiment
seem to be on the verge of illuminating
it~\cite{haule_2009a,chandra_2009a,santander-syro_2009a,elgazzar_2009a}.

The other mysterious phase which was described in the meeting was that
seen in the vicinity of the metamagnetic quantum critical end-point in
${\rm Sr_3Ru_2O_7}$~\cite{grigera_2004b,borzi_2007a}. The latest
results presented by Andy Mackenzie demonstrate that this is a true
phase surrounded by thermodynamic phase transitions with an intriguing
entropy landscape~\cite{rost_2009a,mackenzie-qcnp09}. The suggestion
that it is an electronic nematic phase finds some experimental
support~\cite{borzi_2007a} but is certainly not the only
suggestion. Equally important is the question of what is driving its
formation: is it formed by the quantum critical fluctuations of the
metamagnetic quantum critical end-point, or is the suppression of
metamagnetism simply allowing a sub-leading order instability to
become prominent?

\begin{figure*}[htb]%
  \includegraphics*[width=\textwidth]{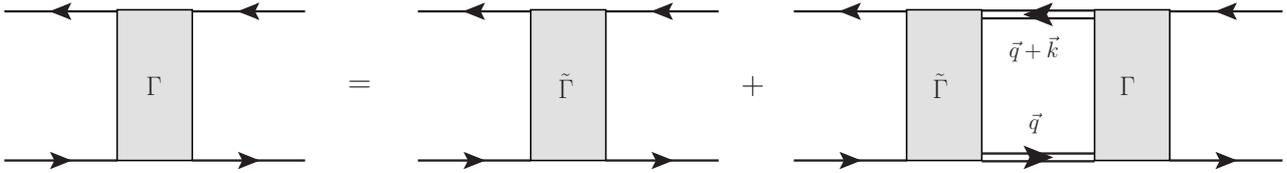}%
  \caption[]{%
By approaching the quantum critical point from within
the Fermi liquid state (along path (b) in Fig.~\ref{qcp}) we can
consider instabilities in the particle-hole channel using the standard
microscopic relations in Fermi liquid theory~\cite{negele_1987a}.  In
particular, the quasiparticle pole in the fully renormalized
single-particle Green's function means that the vertex function
$\Gamma$ can be expressed as in integral equation involving the
non-singular part $\tilde{\Gamma}$ as $\vec{k} \rightarrow 0$. This
can then be used to relate the scattering amplitude to the Landau
parameters and hence the relationship between divergent scattering and
the Pomeranchuk instability.  }
    \label{vertex}
\end{figure*}

How then are we to address these two key issues?  One promising avenue
to investigate both the identity of these ``dark order'' phases and
their origin is to exploit possible parallels with
superconductivity~\cite{schofield_2009a}. Taking our inspiration from
that body of work on spin-mediated Cooper pairing we could ``rotate''
one arm of the pair propagator round [see Fig.~\ref{pairing}(b)]. In
effect we are now asking whether critical fluctuations might cause
``pairing'' singularities in the particle-hole
channel~\cite{lonzarich-old}. These would correspond to exotic types
of magnetism or charge order.

The idea that there is an analogy between superconductivity and
magnetism stems from the very earliest days of BCS
theory~\cite{anderson-1959}. It is also one which has re-emerged in
the intervening years.  Christian Pfleiderer provided us with an
excellent example of this in describing his work on the novel magnetic
texture seen in MnSi~\cite{muhlbauer_2009a}. It seems likely that this
is a skyrmion lattice which can be thought of as the magnetic analogue
of the mixed state of a superconductor~\cite{bogdanov_1989a}.  The
mixed state is a rather conventional superconducting state whose
magnetic analogue is more exotic. What if we now consider more
exotic superconductors and ask what magnetic states they correspond
to? Unconventional superconductors which pair at finite angular
momentum correspond to Pomeranchuk distorted nematic fluids which
break rotational, but not translational, symmetries~\cite{wu_2007a}.
The Fulde-Ferrel-Larkin-Ovchinikov (FFLO)
state~\cite{fulde_1964a,larkin_1964a} state maps to a spiralling
spin-density wave~\cite{berridge_2009a}. Other states like a
$d$-density wave~\cite{chakravarty_2001b} state can also be viewed as
one element of a rather generic mapping (see Table~\ref{analogue}).

Yet our crude rotation from the particle-particle to the particle-hole
channel suggests there could be more to this analogy than simply one
of enumerating possible order parameters. It might be used to indicate
the properties of such phases. For example, we have recently exploited
this mapping to show that the sensitivity to disorder that
characterizes unconventional superconductors is parallelled by a
Pomeranchuk (nematic) phase~\cite{ho_2009a}. 

This analogy may also be able to give insight into the mechanism
behind the formation of such novel magnetic phases, at least in
certain cases. One such case, perhaps relevant for ${\rm Sr_3Ru_2O_7}$, is
whether ferromagnetic fluctuations (like those of metamagnetism)
favour a Pomeranchuk type instability? One imagines approaching the
ferromagnetic quantum critical point from the paramagnetic side at
$T=0$ [{\it i.e.} along path (b) in Fig.~\ref{qcp}]. (We assume
sufficient uniaxial anisotropy to prevent the transition being driven
first order by soft transverse modes~\cite{belitz_2005b}.) Any
putative new phase should develop before the quantum critical
point. It would be an instability in the Fermi liquid state and hence
addressable under the assumptions of Fermi liquid theory. 

Tracing the effect of the fluctuations in the particle-hole channel
means studying vertex function. In the case of the Fermi liquid the
self-consistent equations for the particle-hole vertex function (see
Fig.~\ref{vertex}), are well known~\cite{negele_1987a}. They describe
the connection between the on-shell (scattering) part of the vertex
function and the off-shell Landau parameters
\begin{equation}
B^{i}_l = \frac{F^{i}_l}{1 + \frac{F^{i}_l}{2l+1}} \; .
\end{equation}
Here $l$ is the angular momentum channel and $i=s$ or $a$ for spin
symmetric or antisymmetric components.  This tells us that the
scattering amplitude $B_l^{i}$ in a given angular momentum channel
diverges at the point when the Landau parameter $F^{i}_l$ in the {\em
same} angular momentum channel approaches the Pomeranchuk instability
condition
\begin{equation}
\frac{F_l^i}{2l+1} \rightarrow -1 \;.
\end{equation}

Taken at face value this suggests that say, an instability in the
Stoner channel ($l=0$, $i=a$), is uncoupled from Pomeranchuk
distortions in higher angular momentum channels $(l \neq 0)$ so
ferromagnetic fluctuations do not promote such a distortion. This
contrasts with superconductivity where magnetic (particle-hole)
fluctuations induce particle-particle pairing in other angular
momentum channels. Yet this is not quite the end of the story since
the antisymmetry of the wavefunction couples the different angular
momentum together in the Landau relation $0=\sum_l
(B_l^s+B_l^a)$. Thus to properly answer this question a microscopic
example needs to be studied in more detail. In fact since this meeting
Maslov and Chubukov have independently made progress in this
regard~\cite{maslov-2010a}. Whether or not there is a unified
framework accounting for novel phase formation at quantum critical
points, there is clearly much remaining to be discovered.

\section{Conclusions}
Quantum criticality and associated novel phases represent some of the
most significant challenges to our understanding of condensed matter
systems. The veracity of this was captured by the panel discussion
which was chaired by Piers Coleman and summarized elsewhere in these
Proceedings~\cite{coleman-qcnp09}. Capturing all the activity,
discussion, argument and debate is an impossible task and this summary
is woefully incomplete. However, drawing the threads together, we have
been inspired by new types of materials and even new controllable
experimental systems in our quest to explore these intriguing
phenomena. The continuing surprises at every twist and turn in this
story indicate that nature is clearly more imaginative than we are!
This meeting has also seen the application of new experimental tools
and techniques to investigate these surprises.  The theoretical
questions that are being raised and the new ideas required to answer
them are inspiring this, and a new, generation of scientists to take
up these challenges.

\begin{acknowledgement}
I acknowledge many useful conversations with fellow attendees of the
meeting but I am particularly indebted to conversations with Gil
Lonzarich. I also gratefully acknowledge the continuing support of the
Royal Society and the EPSRC, as well as the organizers of QCNP09.
\end{acknowledgement}

%
%

\providecommand{\WileyBibTextsc}{}
\let\textsc\WileyBibTextsc
\providecommand{\othercit}{}
\providecommand{\jr}[1]{#1}
\providecommand{\etal}{~et~al.}

\end{document}